\documentclass[aps,prd,twocolumn,floatfix,nofootinbib,showpacs,superscriptaddress]{revtex4-1}
\usepackage{amsmath,amsfonts,amssymb,bm}
\usepackage{graphicx}
\usepackage{color}
\usepackage{subfigure}
\usepackage{multirow}
\usepackage{textcomp}
\usepackage{slashed}

\definecolor{purple}{rgb}{0.5,0,0.5}
\definecolor{blue}{rgb}{0.0,0,0.9}
\usepackage[colorlinks=true, pdfstartview=FitV, linkcolor=purple, citecolor= purple, urlcolor=blue]{hyperref}

\begin{document}



\title{Two Photon Transition Form Factor of $\bar{c}c $ Quarkonia}




\author{Jing Chen}
\affiliation{Department of Physics and State Key Laboratory of Nuclear Physics and Technology, Peking University, Beijing 100871, China}
\affiliation{Collaborative Innovation Center of Quantum Matter, Beijing 100871, China}

\author{Minghui Ding}
\affiliation{School of Physics, Nankai University, Tianjin 300071, China}

\author{Lei Chang}
\altaffiliation{Corresponding author: leichang@nankai.edu.cn}
\affiliation{School of Physics, Nankai University, Tianjin 300071, China}

\author{Yu-xin Liu}
\altaffiliation{Corresponding author: yxliu@pku.edu.cn}
\affiliation{Department of Physics and State Key Laboratory of Nuclear Physics and Technology, Peking University, Beijing 100871, China}
\affiliation{Collaborative Innovation Center of Quantum Matter, Beijing 100871, China}
\affiliation{Center for High Energy Physics, Peking University, Beijing 100871, China}

\date{\today}

\begin{abstract}
The two photon transition of $\bar{c}c$ quarkonia are studied within a covariant approach based on the consistent truncation scheme of the quantum chromodynamics Dyson-Schwinger equation for the quark propagator and the Bethe--Salpeter equation for the mesons.
We find the decay widths of $\eta_{c}^{} \to \gamma\gamma$ and $\chi_{c0,2}^{} \to \gamma\gamma$ in good agreement with experimental data.
The obtained transition form factor of $\eta_{c}^{} \to \gamma\gamma^{\ast}$
for a wide range of space-like photon momentum transfer squared
is also in agreement with the experimental findings of the BABAR experiment.
As a by-product, the decay widths of $\eta_{b}^{},\chi_{b0,2}^{} \to \gamma\gamma$ and 
the transition form factor of $\eta_{b}^{}, \chi_{c0,b0}^{} \to\gamma\gamma^{\ast}$
are predicted, which await for experimental test.
\end{abstract}


\maketitle

\section{Introduction}

Charmonium, in a close analogy with positronium, looks like the simplest object for us to understand the quantum chromodynamics (QCD),
the fundamental theory of the strong interaction. However, due to the multi-scales coexistence, both perturbative and non-perturbative features of QCD show up within charmonium system, challenging our understanding of QCD and making it a testing ground for the varianty of theoretical methods
of QCD. Furthermore, the contemporary data suggests very strongly that there exists exotic four-quark states and also charmonium molecules.
Charmonium still presents us new intriguing puzzles though it has been studied for more than 30 years.

In this paper we focus on the two-photon decay widths of charmonia, 
which have attracted theoretical and experimental interests for many years (see, e.g., Refs.~\cite{BaBar2010etac,Pham2013,Belle2013etac,BSEIIIetac2012,CLEO2008chi,BSEIIIetac2012,ZPLi:1991PRD,Gupta:1996PRD,Godfrey:1985PRD,Munz:1996NPA,Ebert:2003MPLA,Huang:1996PRD,Bodwin:1995PRD,Feldmann:1997PLB,Petreli:1998NPB,Lansberg:2006PRD,Lattice-2006,Lattice-2016,Dorokhov:2012NPBPS,Bashir:2016PRD,NNLO12}). For example, the precise data of two-photon decay provides information of the mesons' leptonic decay and their inner parton structure and they are highly relevant for phenomenology. 
On experimental side, considerable progresses on the measurement of $\eta_{c}^{}$ decay have been made from Belle, Babar, CLEO-c and BES (see, e.g., Refs.~\cite{BaBar2010etac,Pham2013,Belle2013etac,BSEIIIetac2012}).
The $\chi_{c0,2}^{}$ decay, which was previously measured by CLEO-c~\cite{CLEO2008chi}, is also reported by BESIII with highly precision data~\cite{BSEIIIetac2012}.

On theoretical side, these decay channels have been extensively studied by various theoretical methods (see, e.g., Refs.~\cite{Godfrey:1985PRD,ZPLi:1991PRD,Bodwin:1995PRD,Gupta:1996PRD,Munz:1996NPA,Huang:1996PRD,Feldmann:1997PLB,Petreli:1998NPB,Ebert:2003MPLA,Lansberg:2006PRD,Lattice-2006,Lattice-2016,Dorokhov:2012NPBPS,Bashir:2016PRD,NNLO12}). Among these calculations, the decay width of pseudoscalar meson given by Lattice QCD simulations~\cite{Lattice-2006,Lattice-2016} is smaller than the current experimental data.
The reason is still not clear.
The non-relativistic QCD (NRQCD) has been widely used to study charmonium physics.
However, the recent calculation including the next-to-next-to-leading-order (NNLO) perturbative correction~\cite{NNLO12} provides large unpleasant deviation in contrast to the leading order result, challenging the `systmematically improved' philosophy of NRQCD.

A natural framework for studing bound state problem in QCD is provided by the Dyson-Schwinger equations (DSEs) \cite{DSErev1,DSErev2,DSErev3,DSErev4}
for the quark gap equation and the
Bethe-Salpeter equation (BSE).
DSEs have been extensively implemented
to study the light hardon physics~\cite{DSErev3,DSErev4}, QCD phase transition~\cite{DSEphase1,DSEphase2,DSEphase3,DSEphase4} and other interesting aspects~\cite{DSErev4}.
The previous application of this approach to two photon decays of heavy quarkonia can be tracked to nearly 20 years ago~\cite{DSEheavy1,DSEheavy2}. There the authors performed a model expression of the quark-photon interaction vertex as input within a consistent truncation of quarks and mesons.
The similar framework with improved expression of quark-photon vertex has been used by Maris and Tandy to study the light hadron system, such as the electromagnetic properties~\cite{MT1} and strong decays~\cite{MT2}. 
As these practitioners find, there exist a series of complex-valued singularities with the increasing of photon momentum square in the numerical Euclidean momentum integration, which limits the calculation within a small domain of photon momentum.
To overcome this weakness, a novel method, with the perturbation theory integral representations (PTIRs) of quark propagator, meson amplitude and quark-photon vertex, has been implemented to calculate the pseudoscalar meson electromagnetic form factor~\cite{LCPionEMFF} and transition form factors~\cite{KhepaniPionTFF,KhepaniEtaTFF} to any space-like momenta. This method provides a unified framework to understand the relation between form factors and meson inner valence quark structures, which can be well extended to the computation of other form factors without a doubt.

In the practical implementation of PTIRs method an additional parameter introduced to fit the decay width~\cite{KhepaniPionTFF,KhepaniEtaTFF}, taking different values for different mesons,
does not truely affect the modest and high photon momentum dependence of the form factors 
but a small domain of momentum, such as that around the interaction radius.
Up to now no one has yet completed any calculation for the heavy quarkonia two photon decay channels within a symmetry-preserving framework of the DSEs. We investigate then the charmonium two photon decay and transition form factor via the similar method as Maris and Tandy have taken in this paper.
We also extend our calculations to predict the bottomonium two photon transition properties. 
The development of this contribution provides a valued complementation for the DSEs PTIRs approach.

The remainder of this paper is organized as follows.
In section~\ref{framework} we describe firstly the framework for calculating the two photon decay widths and the transition form factor of $\eta_{c}^{}$ and $\chi_{c0,2}^{}$ in a covariant approach and then test our numerical technique by calculating the $\pi-\gamma\gamma^{\ast}$ transition form factor.
In section~\ref{Results} we represent our main numerical results for the charmonium decay widths and the transition form factors,
and display our predictions for the bottomonium two photon transition properties.
In section~\ref{Summary} we give our summary and remarks.
An appendix is added to show the covariants for the mesons with $J^{PC}=0^{-+},0^{++},2^{++}$ and the quark-photon vertex in our calculations.

\section{Calculation Framework}
\label{framework}

\subsection{Truncation}

\begin{figure}[ht!]
\centering
 \includegraphics[width=0.32\textwidth]{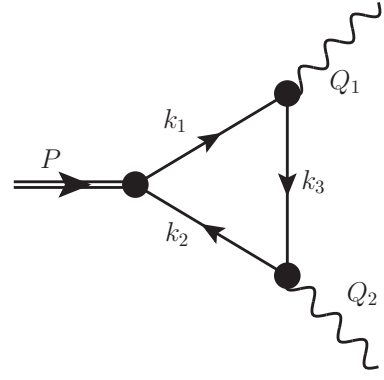}
 \caption{\label{fig:TFF0F} Sketch of the general impulse approximation of meson transition form factor}
\end{figure}

Herein we take the Euclidean metric where $\{\gamma_{\mu},\gamma_{\nu}\}=2\delta_{\mu\nu}$, $\gamma_{\mu}^{\dag}=\gamma_{\mu}$ and $a\cdot b=\sum_{i=1}^{4}=a_{i}b_{i}$.
The general impulse approximation for the transition amplitude of $M^{\bar{q}q}\rightarrow\gamma^{\ast}\gamma$, illustrated in Fig.~\ref{fig:TFF0F}, can be expressed as
\begin{eqnarray}\nonumber
&& \Lambda^q_{\alpha\beta}(Q^2_1,Q^2_2) = e^2 N_c \int^\Lambda_{d k} \text{tr} \left[ S(k_1) \Gamma^{q\bar{q}}_M(k;P) S(k_2) \right.\\\label{eq:transitionamplitude}
&& \qquad \left. \times i\Gamma_\beta(k+Q_1/2;Q_2) S(k_3) i\Gamma_\alpha(k-Q_2/2;Q_1)\right], \quad 
\end{eqnarray}
where $k$ and $P$ are the relative and total momentum of the meson, $Q_1$ and $Q_2$ are the momenta of the photons. $k_1=k+P/2,\;k_2=k-P/2,\;k_3=k+(Q_1-Q_2)/2$ are the momenta of the inter quark with flavor index $q$. tr represents the trace over the Dirac index.
The notation $\int^\Lambda_{d k}=\int ^{\Lambda} d^{4} k/(2\pi)^{4}$ stands for a Poincar$\acute{\text{e}}$ invariant regularized integration, with $\Lambda$ the regularization mass-scale. The regularization can be removed at the end of all calculations by taking the limit $\Lambda\to\infty$.

The transition amplitude of $\pi^0$, $\eta_{c}^{}$ and $\chi_{c0,2}^{}$ can be expressed by multiplying Eq.~(\ref{eq:transitionamplitude}) a charge constant as
$$\Lambda^{\pi^0\to\gamma\gamma^{\star}}_{\alpha\beta}(Q_1;Q_2)
=Z_{\pi^0}\Lambda^{u}_{\alpha\beta}(Q_1;Q_2) \, , $$
$$\Lambda^{\eta_c/\chi_{c0/2}\to\gamma\gamma^{\star}}_{\alpha\beta}(Q_1;Q_2)
=Z_{\eta_c/\chi_{c0/2}}\Lambda^c_{\alpha\beta}(Q_1;Q_2)\, , $$
with
$Z_{\pi^0}=\frac{2}{\sqrt{2}}((\hat{Q}^u)^2-(\hat{Q}^d)^2)=\frac{\sqrt{2}}{3}$ and $Z_{\eta_c}=Z_{\chi_{c0}}=Z_{\chi_{c2}}=2(\hat{Q}^c)^2=8/9$.
We have considered the photon cross term contribution in the factors $Z_{\pi^0,\eta_{c}^{},\chi_{c0/2}^{}}$.
In the discussion of $\pi^{0}$ case we take the isospin symmetry approximation in which $u$- and $d$-quarks have identical strong interactions and properties except electric charge.
The amplitude is only complete when the quark propagator $S$, meson Bethe-Salpeter amplitude (BSA)  $\Gamma^{q\bar{q}}_M$ and the dressed quark-photon interaction vertex $\Gamma_\alpha$ have been fully specified, which will be described in the follows.

In Euclidean space the matrix element for the transition form factor of pseudoscalar meson $0^{-+}$ can be written as
\begin{eqnarray}
 \Lambda^{{\pi^{0}}/{\eta_{c}^{}} \to\gamma\gamma^{\star}}_{\alpha\beta}(Q^{2}_{1},Q^{2}_{2}) 
 & = &  \frac{2i {\alpha_{em}^{}}}{\pi f_{\pi/{\eta_{c}^{}}}^{}} \epsilon_{\alpha\beta\mu\nu}^{} Q_{1\mu}Q_{2\nu} \nonumber \\
  & &  \times F_{{\pi^{0}}/{\eta_{c}^{}} \to\gamma\gamma^{\star}}(Q^{2}_{1},Q^{2}_{2}) \, , 
\end{eqnarray}
where $\alpha_{em}^{}=1/137$ is the fine structure constant, 
$\epsilon_{\alpha\beta\mu\nu}^{}$ is the Levi-Civita symbol.
One can find easily the decay width as
\begin{equation}
 \Gamma_{{\pi^{0}}/{\eta_{c}^{}} \to \gamma\gamma} = \frac{|F_{{\pi^{0}}/{\eta_{c}^{}} \to\gamma\gamma^{\star}}(0,0)|^{2} \alpha^{2}_{em} m^{3}_{{\pi^{0}}/{\eta_{c}^{}}}}{16\pi^{3}  f^{2}_{{\pi^{0}}/{\eta_{c}^{}}}}.
\end{equation}
The general form of the transition form factor of scalar meson $0^{++}$ can be written as
\begin{equation}
 \Lambda^{{\chi_{c0}^{}} \to\gamma\gamma^{\star}}_{\alpha\beta}(Q^{2}_{1},Q^{2}_{2}) =  e^{2} g^{\bot\bot}_{\alpha\beta} F_{{\chi_{c0}^{}} \to\gamma\gamma^{\star}}(Q^{2}_{1},Q^{2}_{2})\, ,
\end{equation}
where
\begin{eqnarray}\nonumber
 g^{\bot\bot}_{\alpha\beta} &=& \delta_{\alpha\beta}^{} - Q_{1\alpha}Q_{1\beta}\frac{Q^{2}_{2}}{D} 
 - Q_{2\alpha}Q_{2\beta}\frac{Q^{2}_{1}}{D}\\
 && + (Q_{1\alpha}Q_{2\beta} +Q_{2\alpha}Q_{1\beta})\frac{Q_{1} \cdot Q_{2}}{D},
\end{eqnarray}
with
\begin{equation}
 D = Q^{2}_{1} Q^{2}_{2} - (Q_{1} \cdot Q_{2})^2\, .
\end{equation}
The corresponding decay width reads
\begin{equation}
 \Gamma_{{\chi_{c0}^{}} \to\gamma\gamma} = \frac{ \pi \alpha^{2}_{em}}{m_{{\chi_{c0}^{}}}} 
 \big{|} F_{{\chi_{c0}^{}} \to\gamma\gamma^{\star}}(0,0) \big{|}^{2} \, .
\end{equation}
The general form of the $\chi_{c2}^{}$ transition amplitude is comber-some and we just quote its form at $Q_{1}^{2}=Q_{2}^{2}=0$ which can be decomposed as
\begin{eqnarray}
 \Lambda^{{\chi_{c2}^{}} \to\gamma\gamma}_{\alpha\beta,\mu\nu}(0,0) & = &  e^{2} \big{[} S^{0}_{\alpha\beta,\mu\nu} F^{0}_{\alpha\beta,\mu\nu}(0,0)  \nonumber \\
 & & \quad +S^{2}_{\alpha\beta,\mu\nu}F^{2}_{\alpha\beta,\mu\nu}(0,0) \big{]} \, ,
\end{eqnarray}
where $\mu,\nu$ are the indices of the tensor meson. The first and second part represent the states with helicity $H=0$ and $H=2$ respectively. The structure functions $S^{0,2}$ take the forms
\begin{eqnarray}
 S^{0}_{\alpha\beta,\mu\nu} &=& \Big{(} \frac{q_{\bot\alpha}^{} q_{\bot\beta}^{} }{q_{\bot}^{2}} -\frac{1}{3}g^{\bot}_{\alpha\beta} \Big{)} g^{\bot\bot}_{\mu\nu}, \\
 S^{2}_{\alpha\beta,\mu\nu} &=& g^{\bot\bot}_{\alpha\mu}g^{\bot\bot}_{\beta\nu}  +g^{\bot\bot}_{\alpha\nu}g^{\bot\bot}_{\beta\mu} -g^{\bot\bot}_{\alpha\beta}g^{\bot\bot}_{\mu\nu},
\end{eqnarray}
where
\begin{equation}
 g^{\bot}_{\mu\nu} = \delta_{\mu\nu}^{} - \frac{P_{\mu} P_{\nu}}{P^2}, \quad q_{\bot\mu}^{}=g^{\bot}_{\mu\nu} Q_{1\nu} \, .
\end{equation}
The corresponding decay width is
\begin{equation}
 \Gamma_{{\chi_{c2}^{}}\to\gamma\gamma} = \frac{ 4\pi \alpha^2_{em}}{5m_{{\chi_{c2}^{}}}} \Big{[} \frac{1}{6}|F^0_{\alpha\beta,\mu\nu}(0,0)|^2 + |F^2_{\alpha\beta,\mu\nu}(0,0)|^{2} \Big{]}.
\end{equation}

We are now in the position to discuss the components ($S$, $\Gamma^{q\bar{q}}_M$ and $\Gamma_{\alpha}$) of the impulse approximation transition amplitude. A systematic study demonstrated in Ref.~\cite{Bando1994} indicates that the gauge invariance of impulse approximation of 3-point Green function is true and only true
when the quark propagator, the meson Bethe-Salpeter amplitude and quark-photon interaction vertex satisfy a mutual consistency, i.e., one should define $S$ by the rainbow truncation of the quark Dyson-Schwinger equation and $(\Gamma^{q\bar{q}}_{M}, \; \Gamma_{\alpha})$ by the ladder truncation of the Bethe-Salpeter equation. This consistent framework guarantees the low energy theorems,
in particular, the correct normalization of the form factor of $\pi^{0}$ by the axial anomaly.
However we should note that this scheme can usually not guarantee the gauge invariant for 4-point Green function, such as the process of $\pi$-$\pi$ scattering~\cite{Bicudo:2002PRD},
which motivates the correction in the definition of parton distribution function~\cite{CLPDF}.
The practical improvement of the impulse approximation is systematic but very complicated
which has gone beyond the scope of the present
paper.
We will discuss possible relation between the truncation problem and the interaction width later.

The rainbow truncated DSE for the quark propagator in Euclidean space reads
\begin{eqnarray}\nonumber
S(p)^{-1} & = & Z_{2} i\gamma\cdot p + Z_{4} m_{q}(\mu) + \frac{4}{3}Z^2_2 \int^\Lambda_{d q} \mathcal{G}((p-q)^2) \\
&&  \times \, D^f_{\alpha\beta}(p-q) {\gamma_{\alpha}^{}} S(q) {\gamma_{\beta}^{}} \, ,
\end{eqnarray}
where $D^{f}_{\alpha\beta}(k)=\left(\delta_{\alpha\beta}-\frac{k_{\alpha}k_{\beta}}{k^{2}}\right)\frac{1}{k^{2}}$ represents the free gluon propagator and the effective interaction is denoted by $\mathcal{G}$. 
We make use of the Landau gauge in this paper.
$Z_{2}$ and $Z_{4}$ are the wave function and mass renormalization constant, respectively.
$m_{q}(\mu)$ is the current quark mass at the space-like renormalization point $\mu$.
We perform a  flavor-independent renormalization scheme as explained in Ref.~\cite{CLPDF}
to define the $Z_{2}$ and $Z_{4}$ at the $\mu$.

The meson amplitude will be calculated by solving the homogenous BSE in ladder truncation
\begin{eqnarray}\nonumber
 \Gamma(k;P) \! & = & \! -\frac{4}{3}Z^2_2\int^\Lambda_{d q}  \Big[\mathcal{G}((k-q)^2)  D^{f}_{\alpha\beta}(k-q)  \\\label{eq:homoBSE}
 &&  \times \gamma_{\alpha}^{} S(q_{+})\Gamma(q;P)S(q_{-})\gamma_{\beta}^{} \Big] \, ,
 \end{eqnarray}
where  $k$ and $P$ are the $q\bar{q}$ state's relative and total momenta, respectively,
$q_\pm=q\pm P/2$.
Different types of mesons, such as pseudoscalar, scalar, tensor, etc, are characterized by different Dirac structures.
The most general decomposition for these bound states can be written as
\begin{equation}
  \Gamma^{JP}(k;P) = \sum_{i=1}^{N^{JP}}\tau_{i}^{JP}(k;P)\mathcal{F}_{i}^{JP}(k;P),
 \end{equation}
the index $\{J, P\}$ represent the angular momentum and the $P$-parity of the meson,
$\tau_{i}^{JP}(k;P)$ are the covariants of the BSA
and $\mathcal{F}_{i}^{JP}(k;P)$ the Lorentz scalar coefficients.
The number of the covariants are $N^{0-}=4$, $N^{0+}=4$ and $N^{2+}=8$.
The covariants are listed in Appendix.
We have considered the charge parity and all the scalar coefficients are even function of the quantity $k\cdot P$.
This equation has solutions at discrete values of $P^{2}=-m_{H}^{2}$,
where $m_{H}$ is the meson mass.
The equation determines completely the amplitude $\Gamma(k;P)$
together with the appropriate normalization condition for the bound states.
The meson amplitude is normalized according to the normalization condition~\cite{NakanishiNorm}
\begin{eqnarray}\nonumber
2P_{\mu}^{} & = & \frac{N_c}{N_J}\frac{\partial}{\partial P_\mu}\int^\Lambda_{d q} \text{tr} \big{[} \Gamma(q;-K) \\
&& \times S(q+)\Gamma(q;K)S(q-) \big{]} \big{|}_{P^2=K^2=-m^2},
 \end{eqnarray}
where $N_{c}^{}=3$ is the color number and $N_{J}=2J+1$ is the number of the polarization directions of a meson with angular momentum $J$.
We will later need the following exact expression for the pseudoscalar and vector meson decay constants:
\begin{eqnarray}
f_{0^{-}}P_{\mu} &=& \frac{Z_{2} N_{c}}{\sqrt{2}} \text{tr} \! \int^{\Lambda}_{d k} \!
\gamma_{5}^{} \gamma_{\mu}^{} S(k+)\Gamma_{0^{-}}(k;P)S(k-),  \quad \\
f_{1^{-}}M_{1^{-}} &=& \frac{Z_{2} N_{c}^{}}{3\sqrt{2}} \text{tr} \int^{\Lambda}_{d k} \gamma_{\mu} S(k+)\Gamma^{\mu}_{1^{-}}(k;P)S(k-) .
\end{eqnarray}
%

The general quark-photon interaction vertex is $\bar{\Gamma}_{\mu}^{q} = \hat{Q}^{q}\Gamma_{\mu}^{q}$,
where $\hat{Q}^{q}$ is the $q$-quark electric charge
and the vector vertex $\Gamma_{\mu}^{q}$ satisfies the inhomogeneous BSE under ladder truncation
\begin{eqnarray}\nonumber
\Gamma_{\mu}^{q}(k;P) & = & Z_{2}\gamma_{\mu}^{}
-\frac{4}{3}Z^{2}_{2} \int^\Lambda_{d q} \big{[} \mathcal{G}((k-q)^2) D^{f}_{\alpha\beta}(k-q)  \\\label{eq:inhomoBSE}
 && \times \gamma_{\alpha}^{} S^{q}(q_{+}) \Gamma_{\mu}^{q}(q;P) S^{q}(q_{-}) \gamma_{\beta}^{} \big{]} \, ,
 \end{eqnarray}
where the general decomposition of the vertex is given in Appendix.

These equations are consistent and coupled with an effective coupling function~$\mathcal{G}(s)$~\cite{Chang:2009PRL}, for which we employ the infrared constant Ansatz~\cite{Qin:2011PRC}
\begin{equation}\label{eq:gluonmodel}
 \frac{\mathcal{G}(s)}{s}=\frac{8\pi^2}{\omega^4}D e^{-s/\omega^{2}} +\frac{8\pi^{2} \gamma_{m}^{} \mathcal{F}(s)}{\text{ln}[\tau+(1+s/\Lambda^{2}_{QCD})^2]} \, ,
\end{equation}
where $\mathcal{F}(s)=[1 - \exp(-s/[4m_{t}^{2}])]/s$, $\gamma_{m}^{}=12/(33-2N_{f})$, with $m_{t}=0.5\,$GeV, $\tau=e^{2} - 1$, $N_f=4$, and $\Lambda^{N_{f}=4}_{\text{QCD}}=0.234\,$GeV.
The first term characterized by the parameters $\omega$ and $D$ represents the low- and  intermediate-momentum part of the interaction. The second term describes the ultraviolet part and produces the correct one-loop perturbative QCD limit.
In our calculations, the equations are renormalized at the scale $\mu=2.0\,$GeV.
%

\subsection{Numerical Technique and Test}

With the on-shell condition $P^{2}=-m_{H}^{2}$ and constraints $Q_{1}^{2}=q^{2}$ and $ Q_{2}^{2}=0$,
we parameterize the momenta of the meson and the two external photons explicitly as
\begin{eqnarray}
 P      &=& \left(0, 0, 0, i m_{H} \right),   \\
Q_{1} &=& \Big{(} 0,0,\frac{m_{H}}{2}  +\frac{q^{2}}{2m_{H}},  i\big{(}\frac{q^{2}}{2m_{H}} - \frac{m_{H}}{2} \big{)} \Big{)}, \\
Q_{2} &=& \Big{(} 0,0,-\frac{m_{H}}{2} -\frac{q^{2}}{2m_{H}}, -i\big{(}\frac{q^{2}}{2m_{H}} + \frac{m_{H}}{2} \big{)} \Big{)}  ,  \quad 
\end{eqnarray}
It should also be mentioned that we focus here on the rest frame for the bound state
but keep in mind that the form factors would be frame independent.

Based on the form of photon momenta we have to solve Eq.~(\ref{eq:inhomoBSE})
in a moving frame which has been done exactly in Ref.~\cite{MarisMoving}.
In our present calculation we still solve it in the rest frame and extrapolate it
by analytic continuation.
We make use of the Legendre Polynomial $P_{m}^{}(k^{2})$ for $k^{2}$
and the Chebyshev Polynomial of the second kind
$U_{j}(z)$ for $z$ to extend the $k^{2}$ and $z=\frac{k\cdot P}{\sqrt{k^{2}P^{2}}}$ dependence
to complex plane,
\begin{equation}
 \Gamma(k;P)=\Gamma(k^{2} , z) = \sum_{m=0,j=0}^{N,M}a_{mj}^{} P_{m}^{}(k^2) U_{j}(z) \, ,
\end{equation}
where $a_{mj}^{}$ is the coefficient.
In the calculation $N$ is taken to be equal to the number of the points for the integral
in $k^2$
and we find that $M=2$ for $U_{j}(z)$ is enough for the accuracy.
The numerical uncertainty could be improved systematically.

To test this numerical method we calculate the $\pi^{0}$ transition form factor with the
interaction parameters $\omega=0.5\,$GeV and $D\omega=(0.80\,\text{GeV})^3$,
which produces $m_{\pi}=0.138\,$GeV and $f_{\pi}=0.093\,$GeV.
\begin{figure}[ht!]
\centering
 \includegraphics[width=0.5\textwidth]{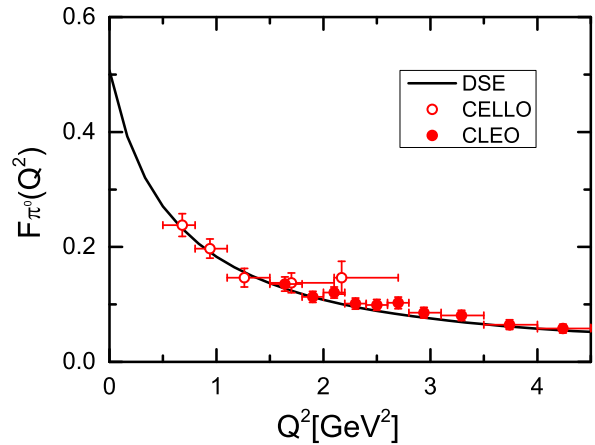}
 \caption{\label{fig:TFFpi} Calculated $\pi^0$ transition form factor within impulse approximation 
 (solid line) and the comparison with experimental data (taken from Ref.~\cite{pionTFFdata}).}
\end{figure}

In Fig.~\ref{fig:TFFpi} we show our calculation results of the $\pi^{0}$ transition form factor $F_{\pi^{0}}(Q^{2})=F_{\pi^{0}\to\gamma\gamma^{\star}}(Q^{2},0)$.
It is evident that the result with our present algorithm agrees with the experiment data
and the previous result given by Maris and Tandy~\cite{MT1} very well,
for instance
$F_{\pi^{0}}(0)=0.507$ obtained in our calculation is almost the same as Maris and Tandy's. 
Such consistence indicates that the transition form factor is not sensitive to the infrared behavior 
of the effective coupling function.

\section{Numerical Results}
\label{Results}

To describe the light quark system within rainbow-ladder truncation it is inevitable to enhance the infrared interaction drastically to give a good understanding of the ground state of light pseudoscalar and vector mesons. On one hand such large enhance is not favored by the solution of gauge-sector gap equation in QCD and Lattice simulation. On the other hand it brings some weaknesses in describing hadron physics,
%
%
for example, the leptonic decay constant of $J/\Psi$ calculated with the same parameters as for light quark system is almost $40\%$ larger comparing to experimental value~\cite{Ding:2016PLB}.

Following Refs.~\cite{Chang:2011PRL,DBkernel}, however, it has become possible to implement
far more sophisticated kernels for the gap equation and Bethe-Salpeter equation,
which overcome the weaknesses of the RL truncation in all channels studied thus far.
This new technique, too, is symmetry preserving; but has an additional strength,
i.e. the capacity to express the dynamical chiral symmetry breaking
(DCSB) nonperturbatively in the integral equations connected with quark-gluon interaction vertex.
Owing to this feature, the new scheme is described as the ``DCSB-improved"
or simply ``DB" truncation.

In a realistic DB truncation the strength of interaction in the infrared region is modest.
Since the DCSB contribution to the dynamical mass of heavy quarks weakens drastically
and in turn all the dressings of the quark-gluon vertex vanishes in the heavy-quark limit,
the RL truncation must become valid (see, e.g., Refs.~\cite{Krassnigg:20115PRD,Ding:2016PLB}).
The modest strength in the infrared region together with the RL truncation would then provide
realistic results in the treatments of heavy-quarkonia.

With $D\omega=(0.70\,\text{GeV})^{3}$ in Eq.~(\ref{eq:gluonmodel}) and $m_{c}^{} (\mu = 2 \, \textrm{GeV}) = 1.835\,\textrm{GeV}$ we produce the mass spectra and decay constants of $\eta_{c}^{}$ and $J/\Psi$ consistent with the Lattice-QCD results~\cite{Becirevic:2014NPB}.
Some of the obtained data are shown explicitly in Table.~\ref{tab:massspectraC}.
Meanwhile we find that the masses and the decay constants of $\eta_{c}^{}$ and $J/\psi$ are almost unchanged with parameter $\omega \in[0.5,0.8]\,\text{GeV}$.
Especially, as $\omega=0.8\,\text{GeV}$, the relative error of the mass spectrum  $|(M^{\textrm{exp}}-M^{\textrm{cal}})/M^{\textrm{exp}}| < 3\%$.
We refer it then as our ``best" parameter.
We would like to mention, in addition, that the interaction favors a larger value of $\omega$ even in the case of chiral limit physics~\cite{Binosi2016} in case beyond the rainbow-ladder truncation.

\begin{table}[h!]
\caption{\label{tab:massspectraC} Calculated mass spectra and decay constants of some $\bar{c}c$ mesons with $D\omega=(0.7\,\textrm{GeV})^3$ (All the masses and decay constants are measured in GeV).}
\begin{tabular}{c|c|c|c|c|c|c|c|c}
\hline \hline
$\;\;\;\omega\;\;\;$&\; $M_{\eta_{c}^{}}\;$	&\; $f_{\eta_{c}^{}}^{}$\;	&  $M_{J/\psi}$ &\;$f_{J/\psi}^{}$ \;&$M_{\chi_{c0}^{}}$&$M_{\chi_{c1}^{}}$&\; $M_{h_{c}^{}}$\;&$M_{\chi_{c2}^{}}$ \\ [0.5mm]
\hline
0.50	& 2.98 	& 0.279 	& 3.13	& 0.298	& 3.29	&3.35	& 3.32	& 3.43 		\\
0.65	& 2.98	& 0.278	& 3.13	& 0.300	&  3.32	& 3.40	& 3.39	& 3.50		\\
0.80	& 2.98  & 0.279	& 3.13	& 0.303	&  3.33 & 3.44	& 3.44	& 3.56		\\
\hline \hline
\end{tabular}
\end{table}

\begin{figure}[t]
\centering
 \includegraphics[width=0.5\textwidth]{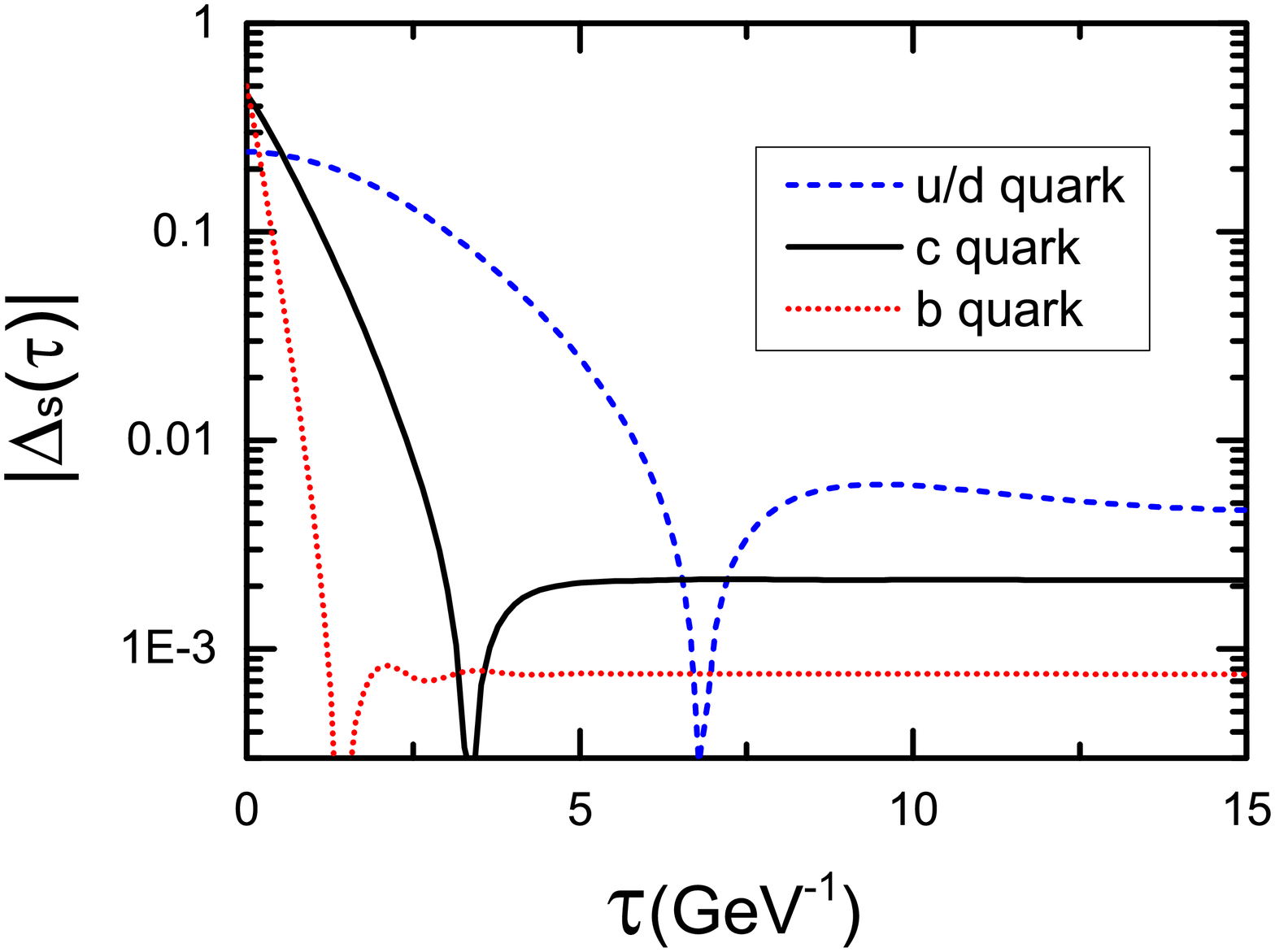}
 \caption{\label{fig:fSchwinger} Calculated results of the Schwinger functions of up/down, charm and bottom quark propagators}
\end{figure}

Furthermore, our calculated charm quark mass function shows nontrivial momentum dependence on the input of the effective interaction.
To explore more information of the quark propagator we calculate the Schwinger function~\cite{DSErev1} with our best parameter,
\begin{equation}
\Delta_{s}(\tau) = \frac{1}{2\pi} \int_{-\infty}^{\infty} d {p_{4}^{}} \text{e}^{-i {p_{4}^{}}\tau} \sigma_{s}({p_{4}^{2}}) \, ,
\end{equation}
where $\sigma_{s}^{}$ is the Dirac-scalar projection of quark propagator.
The obtained $\tau$ dependence of the Schwinger function is shown in Fig.~\ref{fig:fSchwinger}.
To show the current quark mass dependence we also include the up/down and bottom quarks' Schwinger functions in the figure.
The existing node in the quark Schwinger function indicates that the charm and bottom quark propagator as well as the up/down quark propagator take at least a single pair of complex conjugate poles. The poles run away from the origin as the quark mass increases.
Such a behavior of Schwinger function manifests that the spectrum density of the quark is non-positive definite, which supports the idea of confinement of the quark. 
It is also noted that the existing of poles limits the present calculation to high photon momentum when we calculate the transition form factor.

\subsection{Decay Width}

The calculated decay widths of $\eta_{c}^{}$ and $\chi_{c0,c2}^{}$ with several values of the parameter $\omega$ under constraint $\omega\in[0.50,0.80]\,$GeV are listed in Table~\ref{tab:decaywidthC}.
It is straightforward that the inverse of the parameter $\omega$ mimics the effective interaction range (see, e.g., Refs.~\cite{Holl2004,DSEphase1}). Then the $1/{\omega}$ has been used to explore the state's sensitivity to the details of the long-range part of the strong interaction~\cite{Holl2004}.
\begin{table}
\caption{\label{tab:decaywidthC} Calculated two photon decay widths of $\eta_{c}^{}$, $\chi_{c0}^{}$ and $\chi_{c2}^{}$ with several values of $\omega$ under constraint $D\omega=(0.70\, \text{GeV})^3$ (All the decay widths are measured in keV. The world average data of $\eta_{c}^{}$ decay is taken from Ref.~\cite{PDG2016}, and others from Ref.~\cite{BSEIIIetac2012}).}
 \begin{tabular}{c|c|c|c|c}
\hline \hline
 & \, \, $\omega=0.5$\, \,  &\, \,  $\omega=0.65$\, \,  & \, \, $\omega=0.8$\, \, & exp \\
\hline
 $\Gamma^{{\eta_{c}^{}} \to \gamma\gamma}$    & $6.32$ & $6.36$ & $6.39$ &\, 5.1$\pm$0.4\,  \\
 $\Gamma^{{\chi_{c0}^{}} \to \gamma\gamma}$ & $2.06$ & $2.20$ & $2.39$ &\,  2.33$\pm$0.42\, \\
 $\Gamma^{{\chi_{c2}^{}} \to \gamma\gamma}$ & $0.401$ & $0.464$ & $0.655$  &\,  0.63$\pm$0.10\,\\
 $\sigma_{0/2}$                   & $0.0185$ & $0.00813$ & $0.00413$&\,  0.00$\pm$0.04\,  \\ [1mm]
$\frac{\Gamma^{{\chi_{c2}^{}} \to \gamma\gamma}}{\Gamma^{{\chi_{c0}^{}} \to \gamma\gamma}}$ & $0.191$ & $0.211$ & $0.274$ &\,  0.27$\pm$0.06\,
\\ [1mm]
\hline \hline
\end{tabular}
\end{table}
Just as the behavior of the decay constant, the two-photon decay width of $\eta_{c}^{}$ keeps almost the same value which is larger than the world average experimental value, as the parameter is in the domain $\omega\in[0.5,0.8]\, \textrm{GeV}$.
In our calculation we also find an obvious dependence of $\Gamma^{{\eta_{c}^{}} \to \gamma\gamma}$ on the decay constant $f_{\eta_{c}^{}}$.
To demonstrate the relation explicitly,
we carry out a series calculations and obtain a large set of data about the $\Gamma^{{\eta_{c}^{}}\to\gamma\gamma}$ and the $f_{\eta_{c}^{}}$,
with adjusting the value of interaction strength.
After fitting the data we arrive at a relation 
\begin{equation}
\Gamma^{{\eta_{c}^{}} \to\gamma\gamma}=8\pi \left(\frac{2}{3}\right)^{4} \frac{\alpha_{em}^{2}}{m_{\eta_{c}^{}}}\frac{f_{\eta_{c}^{}}^{2}}{(1+\delta)^{2}}\, ,
\end{equation}
where $\delta$ is a parameter.
It is evident that such an expression is consistent with that given in perturbative QCD approach~\cite{Feldmann:1997PLB} and heavy quark spin symmetry estimation~\cite{Lansberg:2006PRD}.
By composing the calculated value of the decay constant $f_{\eta_{c}}$  in the Lattice QCD simulation and the experimental data with the same formula,
Ref.~\cite{Becirevic:2014NPB}  gives a value $\delta = 0.15$, and claims that such a value is too large.
%
%
In our framework we have the value $\delta=0.03$,
which 
%
%
is definitely consistent with what Ref.~\cite{Becirevic:2014NPB} and the references therein expect.
It is also noted that we can produce the present world experimental data if we take the value of the decay constant $f_{\eta_{c}^{}}^{}=0.24\,\textrm{GeV}$.

It is remarkable that a recent Lattice QCD simulation calculation gives the $\eta_{c}^{}$ decay width
as
$\Gamma=1.122(14)\,\textrm{keV}$~\cite{Lattice-2016}.
This value is smaller than the previous quenched Lattice result~\cite{Lattice-2006},
the world average experimental data and our present result.
We find it is quite impossible to accommodate to this result in a large parameter window within our scheme.
It is then desirable to have a more precise and/or direct measurement of this quantity in future experiments to solve this disagreement.

For the P-wave states' two photon decays,
our results of $\Gamma^{{\chi_{c0}^{}} \to \gamma\gamma}$
and $\Gamma^{{\chi_{c2}^{}} \to \gamma\gamma}$
also compares favourably with experiment data.
The calculated ratio of helicity $0$ and helicity $2$ part of the $\Gamma^{{\chi_{c2}^{}} \to  \gamma\gamma}$ with our best parameter is about $\sigma_{0/2}=0.4\%$
which is consistent with the prediction about $0.5\%$ in Ref.~\cite{Barnes1992}.
Meanwhile the $0^{++}$ and $2^{++}$ states' masses and two photon decay widths
show the same $\omega$-dependence as the case of
the axial-vector states $1^{++}$  and $1^{+-}$.

\subsection{Transition Form Factor}

\begin{figure}[htb]
\centering
 \includegraphics[width=0.5\textwidth]{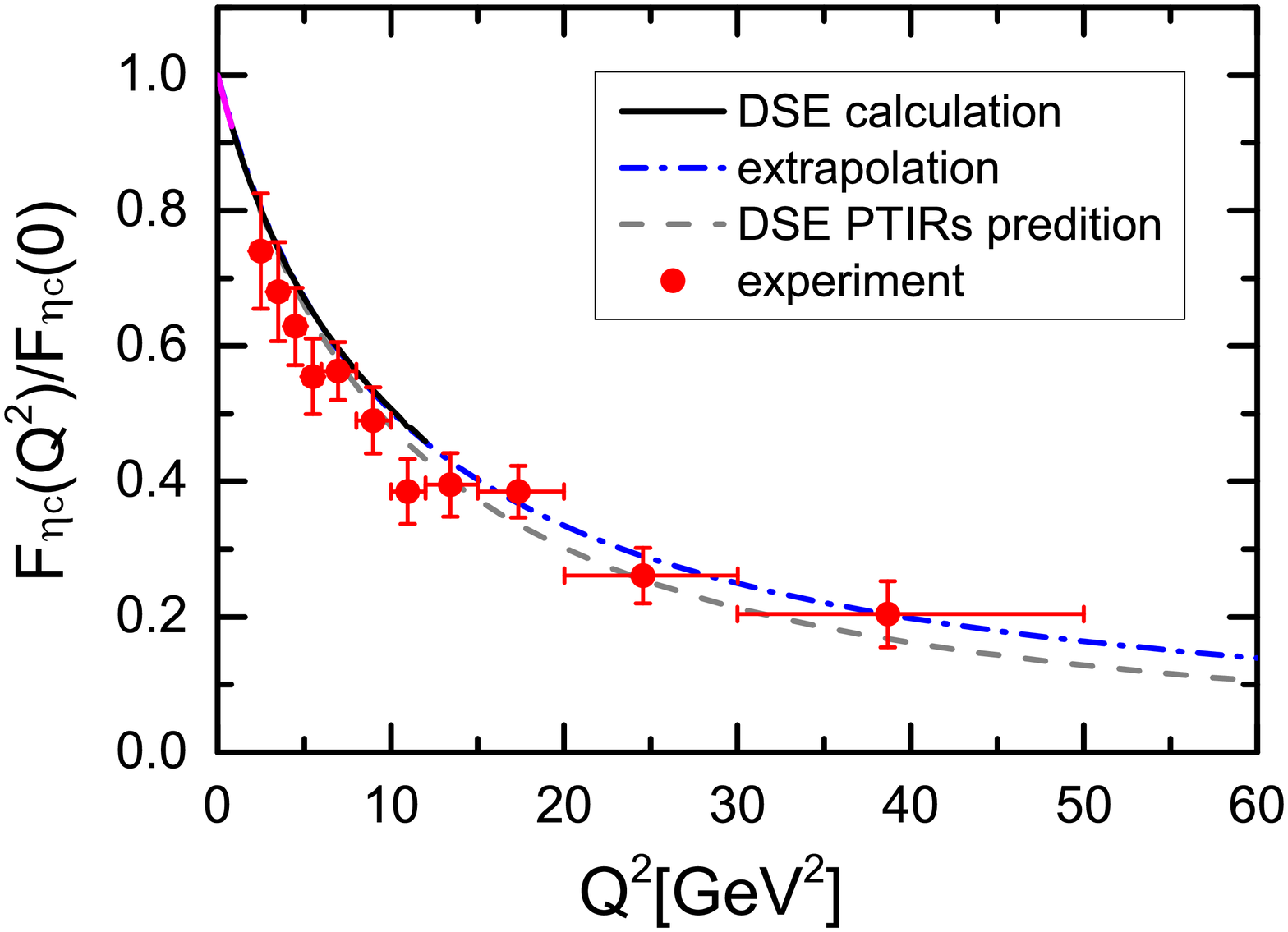}
 \caption{\label{fig:TFFetac} Calculated results of the $\eta_{c}^{}$ transition form factor and the comparison with experimental data.
 Curves: solid line denotes the present impulse approximation calculation result,
 dotted-dashed line is its extrapolation and dashed line presents the DSE PTIRs prediction
 as explained in the text; Data: the experimental data are taken from Ref.~\cite{BaBar2010etac}.}
\end{figure}

The extension of the calculation to finite momentum carried by the virtual photon is straightforward.
In this part we focus on the space-like momentum dependence of the transition form factor.
Our calculated result of the transition form factor $F_{\eta_{c}^{}}(Q^2)=F_{{\eta_{c}^{}}\to\gamma\gamma^{\star}}(Q^2,0)$
is illustrated in Fig.~\ref{fig:TFFetac}.
We can only calculate the transition form factor up to around $12\,\text{GeV}^{2}$ because the
triangular diagram reaches some singularities when $Q^{2}$ becomes larger.
Comparing with the pion case where the domain is only up to  $4\,$GeV, the reliable calculated momentum domain increases as the quark mass increases.
The obtained ratio $F_{\eta_{c}^{}}(Q^2)/F_{\eta_{c}^{}}(0)$ lies in the upper bound of the experimental area and changes very slightly while $\omega \in [0.50,0.80]\,$GeV.
In addition, our result for the $\eta_{c}^{}$ interaction-radius is $r_{\eta_{c}^{}}=0.154\,\text{fm}$,
computed from the slope of the transition form factor.
Such a result agrees with the experimental data $r_{\eta_{c}^{}}= (0.17\pm 0.01)\,\text{fm}$. 
The DSE PTIRs prediction~\cite{KhepaniEtaTFF} has been included in Fig.~\ref{fig:TFFetac}. 
It is noted that the difference between the PTIRs predication and the present direct calculation
is quite slight in the small $Q^{2}$ region.
The deduced interaction radius is almost the same, $r_{\eta_{c}^{}}=0.16\, \text{fm}$, 
in the PTIRs predication. 

As mentioned above, there is a numerical challenge to compute the transition form factor directly on the entire domain of the experimentally accessible momentum transfers illustrated in Fig.~\ref{fig:TFFetac}.
DSE PTIRs suggests a practical way to feature out the transition form factor in the whole momentum space.
Herein we would like to advocate an extrapolation scheme to understand the form factor on the whole domain of space-like momentum.
On one hand we have the impulse approximation results for $F_{\eta_{c}^{}}(Q^{2}=0)$
and the interaction radius $r_{\eta_{c}^{}}$,
on the other hand we have the asymptotic behavior of the impulse approximation
at large $Q^{2}$ as~\cite{BLlimit}
\begin{equation}
\lim_{Q^{2} \to \infty} F_{\eta_{c}^{}}(Q^{2}) = \frac{4\pi^{2} f_{\eta_{c}^{}}^{2}}{Q^{2}} \frac{4}{9} \int_{0}^{1}dx \frac{\varphi_{\eta_{c}^{}}^{}(x)}{1-x} \, ,
\end{equation}
where $\varphi_{\eta_{c}^{}}^{}(x)$ is the leading twist parton distribution amplitude (PDA).
It is noted that the PDA in this expression denotes the asymptotic one
because the meson mass and quark mass scales are not relevant in the large $Q^{2}$ limit.
In general, we have to include the QCD evolution in our approximation to match this limit.
It is tremendously complicated to evolve the diagram from the scale $\mu=2\,\text{GeV}$,
especially considering the fact $\mu\sim m_{H}^{}$.
However the finding in Ref.~\cite{ChangPers} indicates
that the impulse approximation goes to the similar limit with the PDA being fixed at the hadron
scale, for example, $\mu=2\, \text{GeV}$.
We then neglect the possible QCD evolution in our calculation.
We find that the PDA at $\mu=2\,\text{GeV}$ can be expressed as
\begin{equation}
\varphi(\xi) = \mathcal{N} \frac{3}{2} (1-\xi^{2}) \text{e}^{a^{2}((1-\xi^{2})-1)} \, ,
\end{equation}
where $\mathcal{N}$ is the normalized constant, $a=1.6$ for $\omega=0.8\, \text{GeV}$.
It has been known that the distribution amplitude of the heavy quarkonia, which is narrower than the asymptotic form, exhibits a pronounced maximum at $\xi=0$ and is exponentially damped at $\xi \sim \pm 1$ (see, e.g., Refs.~\cite{Dorokhov:2000PLB,Dorokhov:2012NPBPS,Ding:2016PLB}).

Fitting our numerical data in the region $0<Q^{2}<1\,\text{GeV}^{2}$ with constraint from the large $Q^{2}$ limit we can express our transition form factor as
\begin{equation}\label{eq:fitformfatcor}
F(\bar{Q}^{2})/F(Q^{2}=0) = \frac{1+a_{1}^{} \bar{Q}^{2}}{1+b_{1}^{} \bar{Q}^{2}+b_{2}^{} \bar{Q}^{4}}\, ,
\end{equation}
where $Q$ is normalized by $\mu=2\,\text{GeV}$ ($\bar{Q}^{2}=Q^{2}/4$),
%
%
$a_{1}^{}/b_{2}^{}=1.9275$, $b_{1}^{}=0.3908$, and $b_{2}^{}=0.0016$.
The obtained form factor is illustrated as dotted-dashed line in Fig.~\ref{fig:TFFetac}.
We can see that the result in this extrapolation scheme matches our numerical data
in the domain $1\,\text{GeV}^{2}<Q^{2}<12\,\text{GeV}^{2}$ excellently.
Meanwhile this extrapolation can describe the present experimental data well.
The agreement between the DSE PTIRs prediction~\cite{KhepaniEtaTFF} and the impulse approximation at small $Q^{2}$ region is also quite impressive.
At large $Q^{2}$ our extrapolation is just above the PTIRs prediction slightly.
These features indicate that our extrapolation scheme is reliable and easy to carry out in practical calculation.

\begin{figure}[t]
\centering
\includegraphics[width=0.5\textwidth]{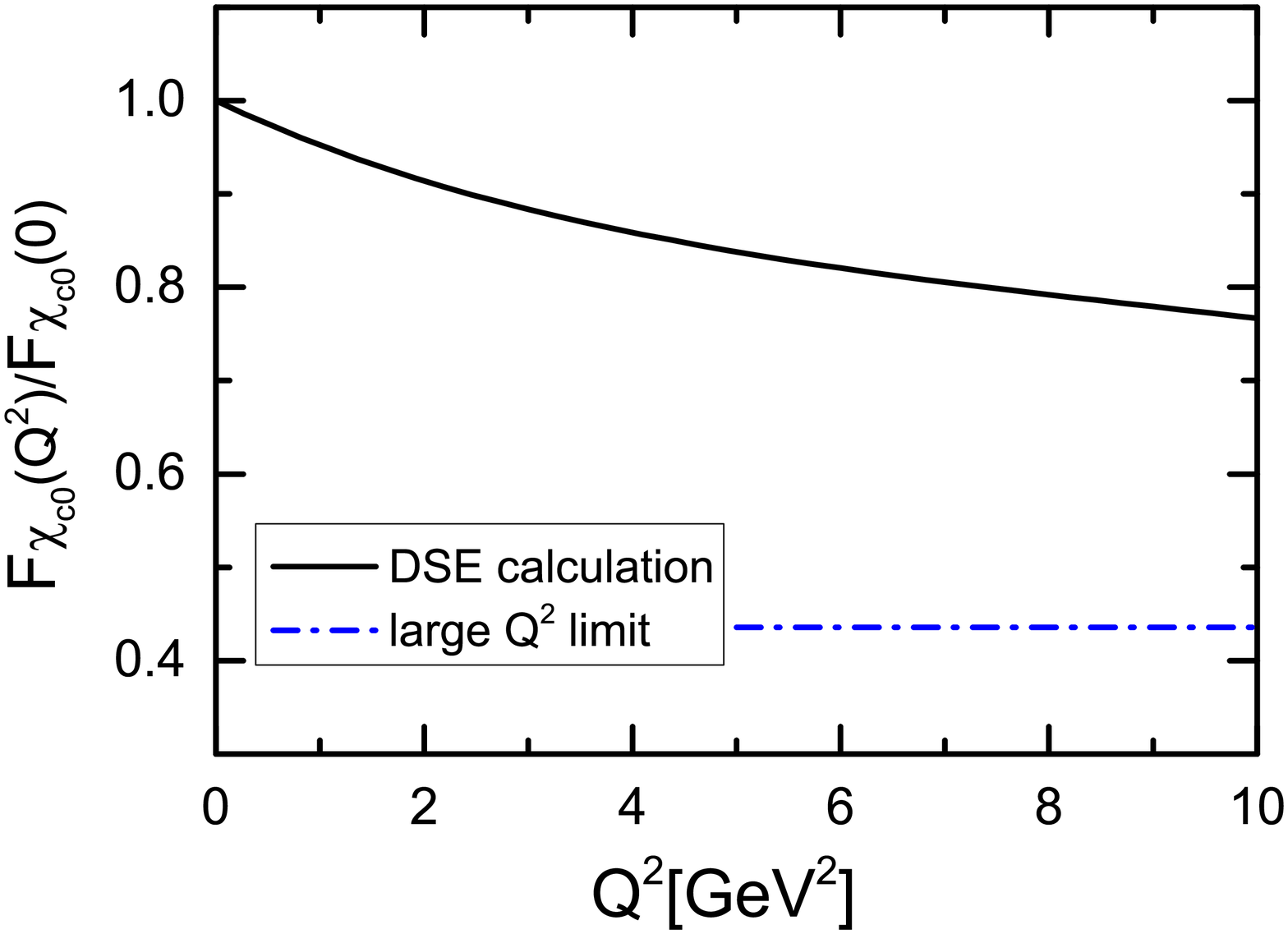}
\caption{\label{fig:TFFxc0} Predication of the $\chi_{c0}^{}$ transition form factor.}
\end{figure}

We also display our result of $\chi_{c0}^{}$ transition form factor $F_{\chi_{c0}^{}}(Q^2)=F_{{\chi_{c0}^{}} \to \gamma\gamma^{\ast}}(Q^2,0)$ in Fig.~\ref{fig:TFFxc0}
although there has no experimental value to compare with.
Therein we also give the result in asymptotic limit as the dashed-dotted line, which is calculated by
\begin{equation}
\lim_{Q^{2}\to\infty} F_{\chi_{c0}^{}}(Q^{2})=\frac{4}{9}\int_{0}^{1}dx \frac{\varphi_{\chi_{c0}^{}}^{}(x)}{1-x}\, ,
\end{equation}
where $\varphi_{\chi_{c0}^{}}^{}(x)$ is the leading twist PDA of $\chi_{c0}^{}$
which has been computed at $\mu=2\,\text{GeV}$ in Ref.~\cite{Ding:2016Prep}.

Following the approach in Ref.~\cite{ChangPers} one can derive such limit exactly within the impulse approximation if one neglects the $\chi_{c0}^{}$ mass.
For the case with the finite mass of $\chi_{c0}^{}$, similar to the case of $\eta_{c}^{}$,
we can not get such limit analytically.
However we can verify that it is true numerically.
We would like to conclude then that the above asymptotic form of $\eta_{c}^{}$ and $\chi_{c0}^{}$
is general where the meson mass is finite and the hadron scale is fixed.

\subsection{Bottomonium}

With the parameters $\omega=0.80\,$GeV and $D\omega=(0.70\,\text{GeV})^3$, we carry out a series calculations to predicate the decay widths of the ground state $\bar{b}b$ meson, the $\eta_{b}^{}$, $\chi_{b0}^{}$ and $\chi_{b2}^{}$.
In these cases, the charge constants are $Z_{\eta_{b}^{}}=Z_{\chi_{b0}^{}}=Z_{\chi_{b2}^{}}=2(\hat{Q}^b)^2 = 2/9$.
The value of the current mass $m_{b}^{}$ of the bottom quark, $m_{b}^{}(\mu=2\,\text{GeV})=7.96\,$GeV, is fixed by the mass of $\eta_{b}^{}$.
Our obtained mass spectra, decay constants and decay widths of the quarkonia are listed in Table~\ref{tab:decaywidthB}.
It is evident that our calculated ratio of the helicity $0$ to helicity $2$ part of $\Gamma^{{\chi_{b2}^{}} \to \gamma\gamma}$, $f_{0/2}^{}\approx 0.0002$, which is just a little bit below the NNLO calculation in NRQCD framework~\cite{NNLO12}.
All these decay widths predication could be tested in future experiments.

\begin{figure}[t]
\centering
 \includegraphics[width=0.5\textwidth]{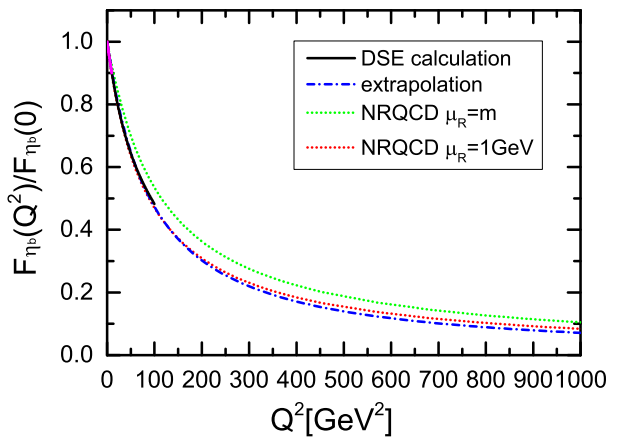}
 \caption{\label{fig:TFFetab} Calculated $\eta_{b}^{}$ transition form factor. Curves: solid line denotes the present impulse approximation calculation result, dotted-dashed line is its extrapolation and dotted lines are NRQCD predictions (green: $\mu_{R}^{}=m$, red: $\mu_{R}^{}=1\,\text{GeV}$).}
\end{figure}

\begin{figure}[t]
\centering
 \includegraphics[width=0.5\textwidth]{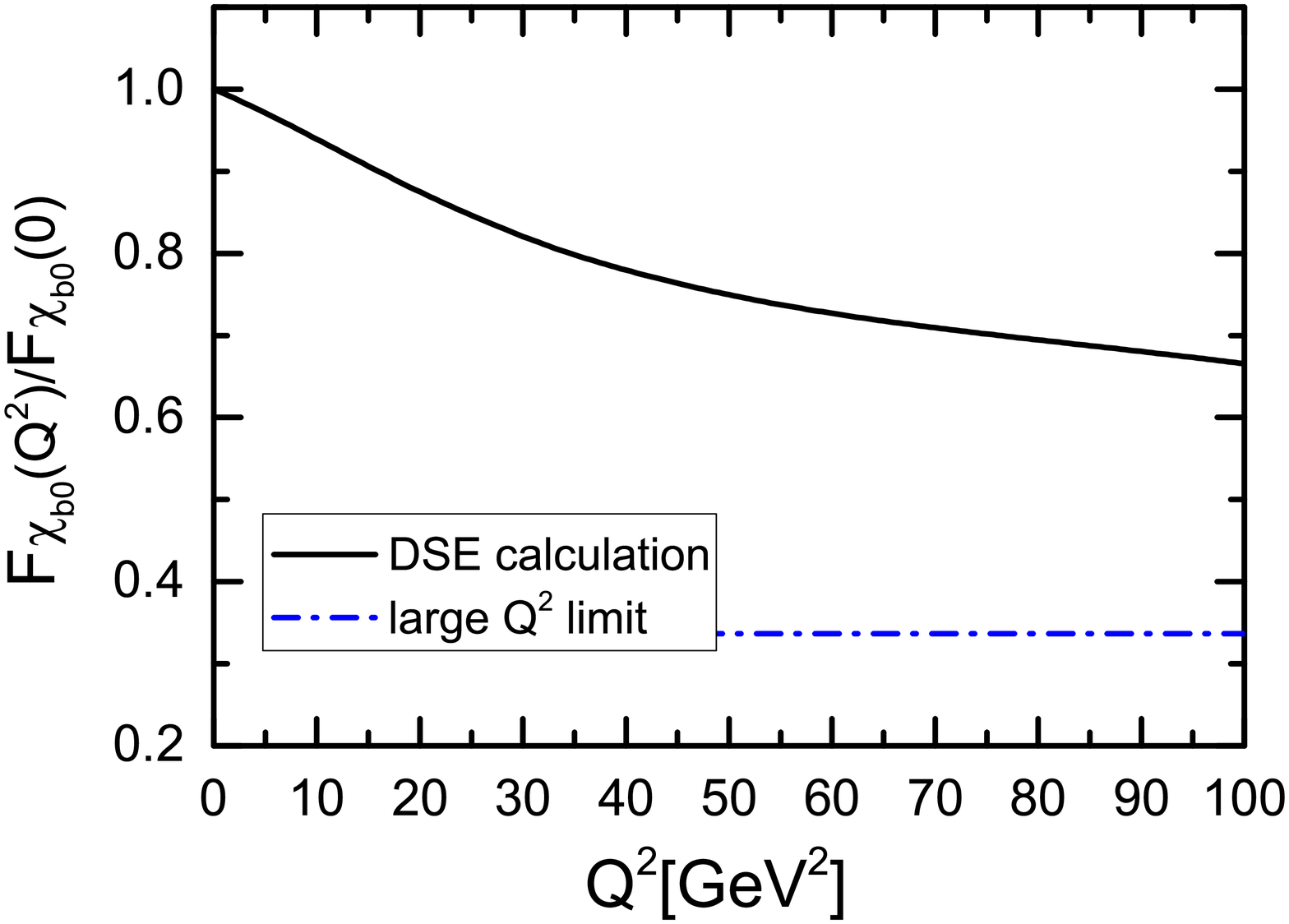}
 \caption{\label{fig:TFFXb0} Predication of the $\chi_{b0}^{}$ transition form factor.}
\end{figure}

\begin{table}[h!]
\caption{\label{tab:decaywidthB} Calculated mass spectra, decay constants and decay widths of some $\bar{b}b$ mesons with $D\omega=(0.70\, \text{GeV})^3$ and $\omega=0.80\,$GeV (All the masses and decay constants are measured in GeV,
and all the decay widths are measured in keV).}
\begin{tabular}{c|c|c|c|c|c|c|c}
\hline \hline
 $M_{\eta_{b}^{}}$	& $f_{\eta_{b}^{}}$	&  $M_{\Upsilon}^{}$ &$f_{\Upsilon}^{}$ &$M_{\chi_{b0}^{}}$&$M_{\chi_{b1}^{}}$& $M_{h_{b}^{}}$	&$M_{\chi_{b2}^{}}$ \\ [0.5mm]
\hline
 9.40  		& 0.509		& 9.47		& 0.491		&  9.81 	& 9.86		& 9.86		& 9.91		\\
\hline
\end{tabular}
\begin{tabular}{c|c|c|c|c}
\hline
$\Gamma^{{\eta_{b}^{}} \to \gamma\gamma}$ 	& $\Gamma^{{\chi_{b0}^{}} \to \gamma\gamma}$
& $\Gamma^{{\chi_{b2}^{}} \to \gamma\gamma}$ & $f_{0/2}^{}$ & $\,\frac{\Gamma^{{\chi_{b2}^{}} \to \gamma\gamma}}{\Gamma^{{\chi_{b0}^{}} \to \gamma\gamma}}$ \\ [0.5mm]
\hline
0.469	&	0.0600	&	0.0143	&	0.000198	&	0.238	\\
\hline \hline
\end{tabular}
\end{table}
Our prediction for $\eta_{b}^{}$ transition form factor $F_{\eta_{b}^{}}(Q^2)=F_{{\eta_{b}^{}}  \to \gamma\gamma^{\ast}}(Q^2,0)$ is shown in Fig.~\ref{fig:TFFetab}.
In the large $Q^2$ limit, it can be written as
\begin{equation}
\lim_{Q^{2}\to\infty} F_{\eta_{b}^{}}(Q^{2}) = \frac{4\pi^{2}f_{\eta_{b}^{}}^{2}}{Q^{2}}\frac{1}{9} \int_{0}^{1}dx \frac{{\varphi_{\eta_{b}^{}}^{}}(x)}{1-x} \, ,
\end{equation}
where $\varphi_{\eta_{b}^{}}^{}(x)$ is the leading twist PDA at $2\,$GeV.
Fitting our data in the region $0< Q^{2} <10\,\text{GeV}$ with the form in Eq.~(\ref{eq:fitformfatcor}), we get the parameters as $a_{1}^{}/b_{2}^{}=16.9$, $b_{1}^{}=0.0480$, $b_{2}^{} = 0.000102$.
Our result matches the NRQCD result (with $\mu_{R}^{}=1\,$GeV) in a large $Q^{2}$ domain well.
The difference in the very large $Q^{2}$ domain could be understood with the value of the decay constant of $\eta_{b}^{}$ and its PDA.
It is worth noting that the NRQCD might be a useful tool for describing bottomonium system
since this effective field theory shows quite good convergence.
This could be attributed to that the current mass of the bottom quark is very large
so that the DCSB plays rather minor role.

Our prediction for the $\chi_{b0}^{}$ transition form factor $F_{\chi_{b0}^{}}(Q^2)=F_{{\chi_{b}^{}} \to \gamma\gamma^{\ast}}(Q^2,0)$ is shown in Fig.~\ref{fig:TFFXb0}.
The large $Q^2$ limit reads
\begin{equation}
\lim_{Q^{2}\to\infty} F_{\chi_{b0}^{}}(Q^{2}) = \frac{1}{9} \int_{0}^{1}dx
\frac{{\varphi_{\chi_{b0}^{}}^{}}(x)}{1-x},
\end{equation}
where $\varphi_{\chi_{b0}^{}}^{}(x)$ is $\chi_{b0}^{}$ leading twist PDA at $\mu=2\,\text{GeV}$.

\section{Summary and Remarks}
\label{Summary}

We completed calculations of the two photon decays of $\bar{c}c$ quarkonia with the impulse approximation, in which all the components are determined by the solutions of the QCD's
Dyson-Schwinger equations (DSEs) and the Bethe-Salpeter equation in the rainbow-ladder truncation, the leading order in a systematic and symmetry-preserving approximation scheme.
The obtained decay widths of the $\eta_{c}^{}$ and the P-wave $\chi_{c0/2}^{}$ are in good agreement
with current experimental data.
We unified the description and explanation of meson--$\gamma\gamma^{\ast}$ transition form factor with the quarkonia $\eta_{c}^{}$ and $\chi_{c0}^{}$, as well as the light pseudoscalar
$\pi^{0}$, via a single DSE interaction kernel.
The constraint on the large momentum limit behavior of the form factor gives us a simple but reliable extrapolation for the $\eta_{c}^{}$ transition form factor.
The prediction of the $\chi_{c0}^{}$ transition form factor within $0-10\,\text{GeV}^{2}$
has also been given in the same framework.
Based on our best parameters we extend our calculations to $\bar{b}b$ quarkonia,
the obtained results await for future experimental test.

This calculation of charmonium two photon decay and transition form factor is based upon the leading order term in a systematic, symmetry-preserving truncation of the equations in quantum field theory.
Quantitative corrections to the results must therefore be expected.
The practical performance beyond the RL truncation~\cite{ChangPDA} indicates
that the quark mass function becomes broader than that in the RL case.
It means that the infrared interaction width becomes smaller
which takes the same meaning as increasing the parameter $\omega$ in the RL truncation.
By considering the $\omega$ dependence of the quantities
we conclude that the two photon decay of $\eta_{c}^{}$ and
its transition form factor are insensitive to the improvement on the truncation.
However we truthfully think that the P-wave states' properties would have a sensitive dependence
on the truncation scheme.
We will leave such a research in the future.

\section*{Acknowledgments}
Work supported by the the National Natural Science Foundation of China with contract No. 11435001
and the National Key Basic Research Program of China with contract
Nos. G2013CB834400 and 2015CB856900 (JC and YXL) and the Thousand Talents Plan for Young Professionals (LC).

\appendix
\setcounter{figure}{0}
\setcounter{table}{0}
\renewcommand{\thefigure}{\Alph{section}.\arabic{figure}}
\renewcommand{\thetable}{\Alph{section}.\arabic{table}}
\section{}
The covariants we use for pseudoscalar, scalar and tensor mesons are
\begin{eqnarray}
\tau_{0^-}^1 &=& i\gamma_5,\\
\tau_{0^-}^2 &=& \gamma_5\slashed{P},\\
\tau_{0^-}^3 &=& \gamma_5\slashed{k}(k\cdot p),\\
\tau_{0^-}^4 &=& i\gamma_5\sigma^{P,k},\\
\tau_{0^+}^1 &=& iI,\\
\tau_{0^+}^2 &=& \slashed{P}(k\cdot p),\\
\tau_{0^+}^3 &=& \slashed{k},\\
\tau_{0^+}^4 &=& i\sigma^{P,k},
\end{eqnarray}
\begin{eqnarray}
\tau_{2^+}^1 &=& M_{\mu\nu},\\
\tau_{2^+}^2 &=& i(M_{\mu\nu}\slashed{k}k\cdot P-2N_{\mu\nu} k\cdot P),\\
\tau_{2^+}^3 &=& iM_{\mu\nu}\slashed{P},\\
\tau_{2^+}^4 &=& M_{\mu\nu}\sigma^{k,P}-2N_{\mu\nu}\slashed{P},\\
\tau_{2^+}^5 &=& iN_{\mu\nu},\\
\tau_{2^+}^6 &=& N_{\mu\nu}\slashed{k},\\
\tau_{2^+}^7 &=& N_{\mu\nu}\slashed{P}k\cdot P,\\
\tau_{2^+}^8 &=& iN_{\mu\nu}\sigma^{k,P}.
\end{eqnarray}
where
\begin{eqnarray}
\sigma^{k,P} &=& \frac{1}{2}(\gamma_\mu\gamma_\nu-\gamma_\nu\gamma_\mu)k_\mu P_\nu,\\
M_{\mu\nu} &=& \gamma^T_\mu k^T_\nu+k^T_\mu \gamma^T_\nu-\frac{2}{3}g^T_{\mu\nu}\gamma\cdot k^T,\\
N_{\mu\nu} &=& k^T_\mu k^T_\nu - \frac{1}{3}g^T_{\mu\nu}k\cdot k^T.
\end{eqnarray}
with $g^{T}_{\mu\nu}=\delta_{\mu\nu}^{} -P_{\mu} P_{\nu}/P^2$, $k^{T}_{\mu} = g^{T}_{\mu\nu}k_{\nu}^{}$.
$k$ and $P$ are the relative and total momenta, respectively.
All the covariants are C-Parity $+$ states.
For the tensor meson, the covariants have two Lorentz index $\mu\nu$.

For the quark-photon interaction vertex there are $12$ independent covariants.
However the longitudinal part and the transverse part can be separated
and only the transverse part contributes to the form factors.
The covariants of the transverse part of the quark-photon vertex we take are
\begin{eqnarray}
\tau_{1^-}^{1} &=& i\gamma^{T}_{\mu} , \\
\tau_{1^-}^{2} &=& i k^{T}_{\mu} \slashed{k},\\
\tau_{1^-}^{3} &=& i k^{T}_{\mu} \slashed{P}(k\cdot p),\\
\tau_{1^-}^{4} &=& \gamma_{5}^{} \epsilon^{T}_{\mu\nu\alpha\beta} \gamma_{\nu}^{} k_{\alpha}^{} P_{\beta}^{}, \qquad \qquad \\
%
\tau_{1^-}^{5} &=& k^{T}_{\mu} , \\
\tau_{1^-}^{6} &=& \sigma^{T}_{\mu\nu} k_{\nu}^{} (k\cdot p) , \\
\tau_{1^-}^{7} &=& \sigma^{T}_{\mu\nu} P_{\nu}^{} ,\\
\tau_{1^-}^{8} &=& k^{T}_{\mu} \sigma^{T}_{\alpha\beta} k_{\alpha}^{} P_{\beta}^{}.  \qquad \qquad
\end{eqnarray}
All the covariants are C-Parity $-$ states.


\end{document}